The manuscript on the following 2 pages is designated for the The manuscript length is limited to 2 pages only.  It was submitted for peer review on 2013-11-04.  Receipt was acknowledged on 2013-11-05 and the manuscript number **TPS7084** was assigned.

# Smoothing of Discharge Inhomogeneities at High Currents in Gasless High Power Impulse Magnetron Sputtering

Joakim Andersson, Pavel Ni and André Anders

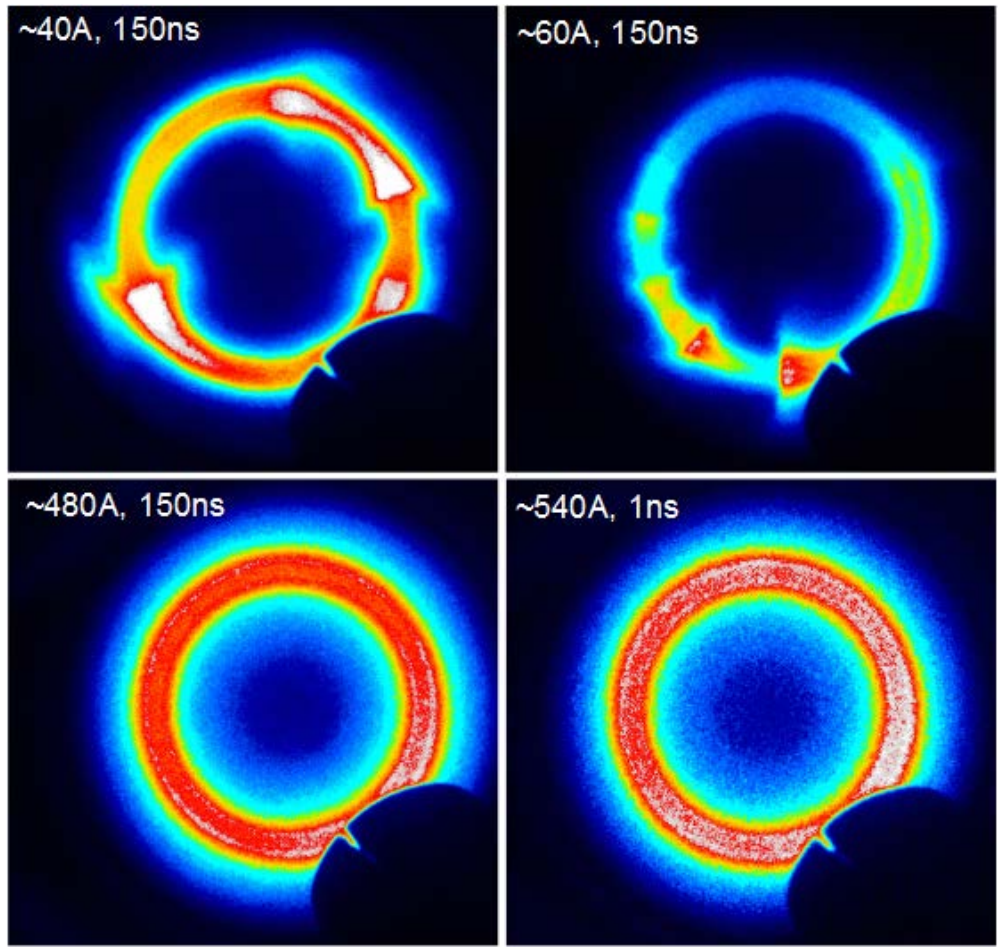

Fig. 1 Four short-exposure images of *gasless* magnetron copper discharges at different currents. At relatively low currents (< 300 A) the discharge shows clear inhomogeneities whereas at higher currents (> 500 A) the discharges are homogeneous. Images are acquired looking at the 76 mm diameter target surface with an image-intensified PI-MAX 1024 camera. For three images the camera was gated to 150 ns to highlight the moving ionization zones. One image is gated to 1 ns to resolve plasma uniformities should they move faster than commonly known. Even at this very short exposure there is no observable azimuthal inhomogeneity above 500 A of discharge current. The cathodic arc plasma source used to initiate the gasless discharge can be seen as dark object in the lower right of each image. Light intensities are presented in false color using the ImageJ look-up table "royal"; the camera was operated in the linear regime, and white does not imply saturation.

*Abstract* - The discharges in high power impulse magnetron sputtering (HiPIMS) have been reported to consist of azimuthally inhomogeneous plasma with locally increased light emission. The luminous zones seemingly travel around the racetrack and are implicated in generation of the high ion kinetic energies observed in HiPIMS. We show that the inhomogeneities smooth out at high discharge current to yield azimuthally homogeneous plasma. This may have implications for the spatial and kinetic energy distribution of sputtered particles, and therefore also on the thin films deposited by high power impulse magnetron sputtering.

Manuscript received … Nov 2013; revised … .
A. Anders and Pavel Ni are with the Lawrence Berkeley National Laboratory, 1 Cyclotron Road, MS 53, Berkeley, California 94720, USA;
J. Andersson is with the Centre for Quantum Technologies, National University of Singapore, 3 Science Drive 2, 117543 Singapore.
Supported by U.S. Department of Energy under Contract No. DE-AC02-05CH11231 and the NRF and the MOE, Singapore.
Publisher Identifier S XXXX-XXXXXXX-X

High power impulse magnetron sputtering (HiPIMS) is pulsed sputtering at high instantaneous power density but low duty cycle (~1%) with the intention to ionize the sputtered particles. Improved quality coatings of many kinds have been deposited using this method [1].

*Gasless* sputtering is a variety of HiPIMS where no sputtering gas is introduced into the chamber [2]. Instead of using process gas as is the case for conventional sputtering, in gasless sputtering each pulse is initiated by a brief plasma puff from a cathodic arc source and continues by "sustained self-sputtering". Sustained self-sputtering was originally observed more than 30 years ago for copper using initiating gas [3]. There are two main advantages. First, the lack of sputter gas gives a particle flux to the substrate that is free of sputter gas as well as contaminants introduced with it, thereby facilitating ultrapure metal coatings. Second, the ratio of metal ions-to-atoms arriving at the substrate can be controlled by the power of the sputtering discharge to affect the coating microstructure and related properties. These magnetron discharges have so far been demonstrated for high sputter yield materials such as silver, copper and nickel.

Discharge inhomogeneities have been reported for several devices based on $\mathbf{E}\times\mathbf{B}$ drift, among them the sputtering magnetron, and several groups have documented flares, spokes and self-organized zones of increased ionization density [4-6]. Such zones were recently implicated as the reason for the much higher kinetic energy ions, compared to DC-sputtering and what is expected from collisional theory, that have been reported to originate in the HiPIMS discharge. The accelerating mechanism of the ions was related to the electric fields of plasma double layers surrounding the traveling inhomogeneities [7]. There are many reports on plasma inhomogeneities in HiPIMS, and they have been thought to be a normal consequence of the high power density. Surprisingly then, it was shown by de los Arcos *et al.* [8], that the now familiar self-organization of the plasma into zones disappears at higher currents, for their 50 mm copper target at ~50 A, to render the discharge azimuthally homogeneous. This happened in a high current and high impedance regime of the discharge. The high impedance was attributed to large amounts of sputtered neutrals, leading to a reduced degree of ionization in the plasma.

Here we share images of a similar phenomenon, although the smoothing of the inhomogeneities occurs at a much higher current in our experiments, about 500 A. The gasless magnetron discharges were run for 200 μs using a pulsed power supply (SIPP by Melec GmbH) on a 76 mm target unbalanced magnetron at a base pressure of $1\times10^{-4}$ Pa. Images were acquired at different currents by changing the camera delay relative to the increasing current pulse. The inhomogeneities are typically triangle-like with the base perpendicular to the $\mathbf{E}\times\mathbf{B}$ direction. Their number and distribution around the racetrack vary. The ten-fold higher current needed to smooth the inhomogeneities in our experiments compared to those of de los Arcos *et al* [8] is only partially explained by the twice larger target area used in our experiments. It is however well known that the discharge is greatly affected by the magnetic field configuration, both in terms of strength and balancing, of the magnetron.

Being able to run the discharge without gas makes another experiment readily possible – investigating the difference in the discharge when sputter gas is added. When running the discharge in argon at a pressure of 0.5 Pa, the smoothing of the inhomogeneities also occurred, but at a slightly lower current, ~400 A, than when running in the gasless mode (~500 A). This difference suggests the presence of neutrals promotes the smoothing of the inhomogeneities. No distinct transition from structured discharge to homogeneous was observed with changing current. Instead, there is a gradual smoothing of the inhomogeneities, in our case when increasing the current from 400 to 500 A. Measurements of ion kinetic energy could give further clues on the underlying mechanisms.